\documentclass[aps,prx,twocolumn,floatfix,groupedaddress,10pt]{revtex4-2}
\usepackage{amsmath,amsthm,amssymb,amscd}
\usepackage{bm}
\usepackage{here}
\usepackage{color}
\usepackage[dvips]{graphicx}
\usepackage{tikz}
\newtheorem*{theorem}{Theorem}
\begin{document}

\title{Linear position measurements with minimum error-disturbance in each minimum uncertainty state}

\author{Kazuya Okamura}
\email[]{k.okamura.renormalizable@gmail.com}
\affiliation{Research Origin for Dressed Photon, 3-13-19 Moriya-cho Kanagawa-ku, Yokohama, 221-0022, Japan}
\affiliation{Graduate School of Informatics, Nagoya University, Chikusa-ku, Nagoya 464-8601, Japan}


\date{\today}


\begin{abstract}
In quantum theory, measuring process is an important physical process;
it is a quantum description of the interaction between the system of interest and the measuring device.
Error and disturbance are used to quantitatively check the performance of the measurement,
and are defined by using measuring process.
Uncertainty relations are a general term for relations that provide constraints on them,
and actively studied.
However, the true error-disturbance bound for position measurements is not known yet.
Here we concretely construct linear position measurements with minimum error-disturbance
in each minimum uncertainty state.
We focus on an error-disturbance relation (EDR), called the Branciard-Ozawa EDR, for position measurements.
It is based on a quantum root-mean-square (q-rms) error and a q-rms disturbance.
We show the theorem that gives a necessary and sufficient condition for
a linear position measurement to achieve its lower bound in a minimum uncertainty state, and
explicitly give exactly solvable linear position measurements achieving its lower bound in the state.
We then give probability distributions and states after the measurement when using them.
It is expected to construct measurements with minimum error-disturbance
in a broader class of states in the future,
which will lead to a new understanding of quantum limits, including uncertainty relations.

\end{abstract}



\maketitle

\section{Introduction \label{1}}
Quantum measurement theory plays a central role both in quantum foundations and in quantum information science.
As well-known, Heisenberg uncertainty relations give
inevitable limitations on measurements for quantum systems
and are treated as one of monumental achievements in early quantum mechanics.
In recent years, quantum measurement technology is rapidly developing and actively used in various devices.
It is now considered that quantum measurements provide information processing
with new computing methods, like measurement-based quantum computation 
\cite{raussendorf2003measurement,briegel2009measurement,zwerger2012measurement}.

Universally valid reformulation of uncertainty relations
is a currently developing topic from theory
\cite{ozawa2002position,ozawa2003physical,ozawa2003universally,ozawa2004uncertainty,
werner2004uncertainty,branciard2013error,ozawa2014errordisturbance,lu2014improved,busch2013proof,
busch2014heisenberg, busch2014measurement,busch2014colloquium} to experiment
\cite{erhart2012experimental,rozema2012violation,baek2013experimental,sulyok2013violation,
weston2013experimental,kaneda2014experimental,ringbauer2014experimental,sulyok2015experimental,
demirel2016experimental, demirel2019experimental, liu2019experimental, liu2019experimental2}.
The violation of the Heisenberg EDR is crucial in this context.
A model of position measurement violating the Heisenberg EDR,
called the error-free linear measurement \cite{ozawa1988measurement,ozawa1989realization},
has already founded in the debate on the sensitivity limit of the gravitational wave detector,
called the standard quantum limit,
in the 1980s \cite{braginsky1980quantum,caves1980measurement,caves1985defense,yuen1983contractive}.
This fact motivated Ozawa \cite{ozawa2002position,ozawa2003physical,ozawa2003universally,ozawa2004uncertainty}
 to formulate universally valid uncertainty relations.
Several experiments \cite{erhart2012experimental,rozema2012violation,baek2013experimental,sulyok2013violation,
weston2013experimental,kaneda2014experimental,ringbauer2014experimental,sulyok2015experimental,
demirel2016experimental, demirel2019experimental, liu2019experimental, liu2019experimental2}
verify universally valid uncertainty relations and the violation of the Heisenberg EDR.
The Branciard-Ozawa EDR treated in the paper is a universally valid trade-off relation for
the noise-operator based q-rms error and the disturbance-operator based q-rms disturbance.
It is first proved by Branciard \cite{branciard2013error}
for vector (or pure) states, and is extended to mixed states by Ozawa \cite{ozawa2014errordisturbance}.
Both the noise-operator based q-rms error and the disturbance-operator based q-rms disturbance
are defined in Sec.~\ref{2.1}.
The goal of the paper is to concretely construct exactly solvable linear position measurements which
achieve the lower bound of the Branciard-Ozawa EDR in each minimum uncertainty state.
Exactly solvable linear measurements are
systematically studied by Ozawa \cite{ozawa1990quantum,ozawa2013disproving}
and the key to achieving our goal.

In the paper, we consider a one-dimensional non-relativistic single-particle system $\mathbf{S}$,
whose position $Q_1$ and momentum $P_1$ at time $0$
are defined as self-adjoint operators on $\mathcal{H}_\mathbf{S}=L^2(\mathbb{R})$
and satisfy the canonical commutation relation (CCR) $[Q_1,P_1]=i\hbar 1$.
A unit vector $\psi$ in $\mathcal{H}_\mathbf{S}$ is called a \textit{minimum uncertainty state}
(or a Gaussian wave packet)
if it satisfies $\sigma(Q_1)\sigma(P_1)=\hbar/2$,
where $\sigma(Q_1)$ and $\sigma(P_1)$ are standard deviations of $Q_1$ and $P_1$ in $\psi$, respectively.
When the state of $\mathbf{S}$ is a minimum uncertainty state $\psi$,
the noise-operator based q-rms error $\varepsilon(Q_1)$ of $Q_1$
and the disturbance-operator based q-rms disturbance $\eta(P_1)$ of $P_1$ satisfy
\begin{equation} \label{BOinq}
\varepsilon(Q_1)^2\sigma(P_1)^2+\sigma(Q_1)^2\eta(P_1)^2 \geq\hbar^2/4
\end{equation}
for all position measurements. This inequality is the Branciard-Ozawa EDR in the minimum uncertainty state $\psi$.
The main result of the paper is to show the lower bound of Eq.~(\ref{BOinq}) is achievable
by using linear position measurements.
That is to say, we concretely construct linear position measurements satisfying
\begin{equation} \label{BOinq2}
\varepsilon(Q_1)^2\sigma(P_1)^2+\sigma(Q_1)^2\eta(P_1)^2 =\hbar^2/4
\end{equation}
in $\psi$. Therefore, when we consider position measurements in each minimum uncertainty state,
no EDR is tighter than the Branciard-Ozawa EDR.
Showing this statement is the main contribution of the paper.

In order to define (exactly solvable) linear position measurements,
we use a one-dimensional non-relativistic single-particle system $\mathbf{P}$,
the probe part of a measuring apparatus $\mathbf{A}$, whose position $Q_2$ and momentum $P_2$
are defined on $\mathcal{H}_\mathbf{P}=L^2(\mathbb{R})$ and satisfy the CCR, $[Q_2,P_2]=i\hbar 1$.
The measuring interaction of linear position measurements for $\mathbf{S}$ is given by
\begin{equation}
H_{int}=K[\alpha Q_1P_2+\beta  P_1Q_2+\gamma (Q_1P_1-Q_2P_2)], \label{LPM}
\end{equation}
where $K(>0)$ is the coupling constant, and $\alpha$, $\beta$ and $\gamma$ are real numbers.
The famous von Neumann model \cite{von2018mathematical} corresponds to the case where $(\alpha,\beta,\gamma)=(1,0,0)$.
On the other hand, the error-free position measurement 
is the case where $(\alpha,\beta,\gamma)=(2,-2,1)/3\sqrt{3}$.
We believe that the rediscovery of exactly solvable linear measurements in the context of uncertainty relations
is another contribution of the paper.

In Sec.~\ref{2}, measuring process, the noise-operator based q-rms error
and the disturbance-operator based q-rms disturbance are defined.
Linear position measurement is then defined.
In Sec.~\ref{3}, we first present a theorem that gives a necessary and sufficient condition 
for a linear position measurement to satisfy Eq.~(\ref{BOinq2}) in $\psi$.
Next, we give three families of linear position measurements satisfying Eq.~(\ref{BOinq2}) in $\psi$.
We then investigate probability distributions and states after the measurement when using
linear position measurements satisfying Eq.~(\ref{BOinq2}) in $\psi$.
In Sec.~\ref{4}, the results of the paper are examined.
In particular, the use of Gauss's error and the noise-operator base q-rms error is discussed.
In Sec.~\ref{5}, we prove the theorem and show a systematic construction of linear position measurements
satisfying Eq.~(\ref{BOinq2}) in $\psi$. Furthermore, we find probability distributions and
states after the measurement when using them via characteristic functions.

\subsection*{Conventions}
Let $\mathcal{H}$, $\mathcal{K}$ be Hilbert spaces and $\phi$ be a unit vector of $\mathcal{H}$.
When $\phi$ is the state of the system, for every self-adjoint operator $A$ on $\mathcal{H}$,
the mean of $A$ and the standard deviation of $A$ are denoted by
$\langle A \rangle_\phi=\langle \phi|A|\phi \rangle=\langle \phi|A\phi \rangle$ 
and $\sigma(A\Vert\phi)=\sqrt{\langle A^2 \rangle_\phi-\langle A \rangle_\phi^2}$, respectively.
As long as there is no confusion, $\langle A \rangle_\phi$ and $\sigma(A\Vert\phi)$ are abbreviated
as $\langle A \rangle$ and $\sigma(A)$, respectively.
For every linear operator $A$ on $\mathcal{H}$ and $B$ on $\mathcal{K}$,
the tensor product $A\otimes B$ of $A$ and $B$, a linear operator on
$\mathcal{H}\otimes\mathcal{K}$, is abbreviated as $AB$.
In particular, we write $A\otimes 1$ and $1\otimes B$ as $A$ and $B$, respectively, for short.
For every self-adjoint operator $A$ on $\mathcal{H}$, $E^A$ denotes the spectral measure of $A$.
Let $n$ be a natural number, $A_1,\cdots,A_n$ mutual commuting self-adjoint operators on $\mathcal{H}$,
and $\phi$ a unit vector of $\mathcal{H}$.
The joint probability measure $\mu^{A_1,\cdots,A_n}_\phi$ of $A_1,\cdots,A_n$ in $\phi$ is defined by
\begin{equation}
\mu^{A_1,\cdots,A_n}_\phi(J_1\times\cdots\times J_n)=\langle \phi|E^{A_1}(J_1)\cdots E^{A_n}(J_n)\phi \rangle
\end{equation}
for all intervals $J_1,\cdots,J_n$ of $\mathbb{R}$.
The probability density function of $\mu^{A_1,\cdots,A_n}_\phi$ with respect to the Lebesgue measure on $\mathbb{R}^n$
is denoted by $p^{A_1,\cdots,A_n}_\phi(a_1,\cdots,a_n)$ if it exists.

\section{Preliminaries \label{2}}

We consider the one-dimensional non-relativistic single-particle system $\mathbf{S}$.
\textit{Throughout the paper, we suppose that the state $\psi$ of the system $\mathbf{S}$
is a minimum uncertainty state such that $\langle Q_1 \rangle_\psi=q_1$, $\langle P_1 \rangle_\psi=p_1$
and $\sigma(Q_1\Vert\psi)=\sigma_1>0$, i.e.,} 
\begin{equation}
\psi(x)=\sqrt[4]{\dfrac{1}{(2\pi)\sigma_1^2}}e^{-\frac{1}{4\sigma_1^2}(x-q_1)^2+i\frac{p_1}{\hbar} x}
\end{equation}
\textit{in the coordinate representation.}
Then $\psi$ satisfies $\sigma(P_1)=\hbar/(2\sigma_1)=:\hat{\sigma}_1$.

\subsection{Measuring process, q-rms error and q-rms disturbance \label{2.1}}

A measuring process for $\mathbf{S}$ is a 4-tuple
$\mathbb{M}_0=(\mathcal{K},\zeta,M,U)$ of a Hilbert space $\mathcal{K}$,
a unit vector $\zeta$ of $\mathcal{K}$, a self-adjoint operator $M$ on $\mathcal{K}$
and a unitary operator $U$ on $\mathcal{H}_\mathbf{S}\otimes\mathcal{K}$.
Here self-adjoint operators and density operators on $\mathcal{K}$ 
describe observables and states of the probe part $\mathbf{P}_0$ of a measuring appratus $\mathbf{A}_0$, respectively.
$\zeta$ is the state of $\mathbf{P}_0$, and $M$ is the meter observable.
$U$ then describes the measuring interaction between $\mathbf{S}$ and $\mathbf{P}_0$
which turns on at time $0$ and turns off at time $\tau(>0)$.
When using $\mathbb{M}_0$, for every observable $X=:X(0)$ of $\mathbf{S}+\mathbf{P}_0$ at time $0$,
the observable $X(\tau)$ at time $\tau$ is given by
\begin{equation}
X(\tau)=U^{-1}XU.
\end{equation}
Then, the noise-operator based q-rms error of $A$ in a vector state $\phi$ is defined by
\begin{equation}\label{NOBE}
\varepsilon(A,\mathbb{M}_0,\phi) =\sqrt{\langle N(A,\mathbb{M}_0)^2\rangle_{\phi\otimes\zeta}},
\end{equation}
where $N(A,\mathbb{M}_0)=M(\tau)-A(0)$ is the noise operator for $A$. For every observable $B$,
the disturbance-operator based q-rms disturbance of $B$ in $\phi$ is defined by
\begin{equation}\label{DOBD}
\eta(B,\mathbb{M}_0,\phi) = \sqrt{\langle D(B,\mathbb{M}_0)^2\rangle_{\phi\otimes\zeta}},
\end{equation}
where $D(B,\mathbb{M}_0)=B(\tau)-B(0)$ is the  disturbance operator of $B$.
Unless confusion arises, $\varepsilon(A,\mathbb{M}_0,\phi)$ and $\eta(B,\mathbb{M}_0,\phi)$
are abbreviated as $\varepsilon(A)$ and $\eta(B)$, respectively.
For any density operator $\rho$ on $\mathcal{H}_\mathbb{S}$, the noise-operator based q-rms error 
$\varepsilon(A,\mathbb{M}_0,\rho)$ and the disturbance-operator based q-rms disturbance
$\eta(B,\mathbb{M}_0,\rho)$ are defined by replacing
$\langle \cdots\rangle_{\phi\otimes\zeta}$ by $\mathrm{Tr}[(\cdots)(\rho\otimes|\zeta\rangle\langle\zeta|)]$
in Eqs.~(\ref{NOBE}) and (\ref{DOBD}), respectively.

We say that two observables $A$ and $B$ are commuting in a vector state $\phi$ if
$[E^A(J_1), E^B(J_2)]\phi=0$ for all intervals(, more generally, Borel sets) $J_1,J_2$ of $\mathbb{R}$.
If two observables $A$ and $B$ are commuting in $\phi$,
then there exists a probability measure $\mu^{A,B}_{\phi}$ such that
\begin{equation}
\mu^{A,B}_{\phi}(J_1\times J_2)=\langle \phi| E^A(J_1)E^B(J_2) \phi\rangle.
\end{equation}
We refer the reader to \cite{ozawa2019soundness}
and references therein for the general treatment of the state-dependent commutativity.

Let $A$ be an observable of $\mathbf{S}$ and $\mathbb{M}_0=(\mathcal{K},\zeta,M,U)$ a measuring process for $\mathbf{S}$.
If $A(0)$ and $M$ are commuting in $\phi\otimes\zeta$, then
the noise-operator based q-rms error $\varepsilon(A,\mathbb{M}_0,\phi)$ satisfies
\begin{equation}
\varepsilon(A,\mathbb{M}_0,\phi) =\varepsilon_G(\mu^{A(0),M(\tau)}_{\phi\otimes\zeta}),
\end{equation}
where Gauss' (rms) error $\varepsilon_G(\mu)$ for a probability measure $\mu$ on $\mathbb{R}^2$ is defined by
\begin{equation}
\varepsilon_G(\mu)=\left(\int_{\mathbb{R}^2} (x-y)^2\;d\mu(x,y) \right)^{\frac{1}{2}}.
\end{equation}

A measuring process $\mathbb{M}_0$ for $\mathbf{S}$ is called a position measurement (for $\mathbf{S}$)
if it is used to measure the position $Q_1$ of $\mathbf{S}$.
Every position measurement $\mathbb{M}_0$ satisfies Eq.~(\ref{BOinq}) in $\psi$. 
In the paper, a position measurement for $\mathbf{S}$ is said to have the minimum error-disturbace in $\psi$
if it satisfies Eq.~(\ref{BOinq2}) in $\psi$.

We omit here the introduction of completely positive instrument, which is a central concept in quantum measurement theory
for describing state changes due to measurements.
We refer the reader to \cite{ozawa2004uncertainty,ozawa1984quantum,okamura2016measurement} for details.
Before stating the main results of the paper, we shall define linear position measurements for $\mathbf{S}$.

\subsection{Linear position measurements \label{2.2}}
As mentioned in Sec.~\ref{1}, we use the system $\mathbf{P}$ to define linear position measurements.
\textit{In considering linear position measurements,
we ignore the intrinsic dynamics of $\mathbf{S}$ and $\mathbf{P}$.}
The composite system $\mathbf{S}+\mathbf{P}$ of $\mathbf{S}$ and $\mathbf{P}$ is described by
the tensor product Hilbert space $\mathcal{H}_{\mathbf{S}}\otimes \mathcal{H}_{\mathbf{P}}\cong L^2(\mathbb{R}^2)$.
The time evolution of the composite system $\mathbf{S}+\mathbf{P}$, the measuring interaction, is described by
the unitary operator
\begin{equation}
U(t)=e^{-\frac{it}{\hbar}H_{int}},
\hspace{5mm}t\in\mathbb{R},
\end{equation}
on $\mathcal{H}_{\mathbf{S}}\otimes \mathcal{H}_{\mathbf{P}}$.
Here the interaction Hamiltonian $H_{int}$ is given by Eq.~(\ref{LPM}).
\textit{Since we ignore the intrinsic dynamics of $\mathbf{S}$ and $\mathbf{P}$,
$K$ contributes only to the time scale of the measurement time. For simplicity, we assume $K=1$ in the paper.}
Each observable $X(t)$ of $\mathbf{S}+\mathbf{P}$ at time $t$
with the initial condition $X(0)=X$ is then given by
\begin{equation}
X(t) = U(t)^{-1}X U(t).
\end{equation}
By solving Heisenberg's equations of motions, we see that
$Q_1(t)$, $Q_2(t)$, $P_1(t)$ and $P_2(t)$ satisfy the following relations:
\begin{subequations}
\begin{align}
\left(
\begin{array}{c}
Q_1(t)  \\
Q_2(t)
\end{array}
\right) &= e^{tS}\left(
\begin{array}{c}
Q_1(0)  \\
Q_2(0)
\end{array}
\right), \\
\left(
\begin{array}{c}
P_1(t)  \\
P_2(t)
\end{array}
\right) &= e^{-tS^T}\left(
\begin{array}{c}
P_1(0)  \\
P_2(0)
\end{array}
\right)
\end{align}
\end{subequations}
for all $t\in\mathbb{R}$, where
\begin{equation}
S=\left(
\begin{array}{cc}
\gamma & \beta \\
\alpha & -\gamma
\end{array}
\right)
\end{equation}
and $S^T$ is the transpose of $S$.

We call a 4-tuple $\mathbb{M}=(\mathcal{H}_\mathbf{P},\xi,Q_2,U(\tau))$ a linear position measurement
for $\mathcal{H}_\mathbf{S}$ (or for $\mathbf{S}$) if we use $Q_2$ to measure $Q_1$,
where $\tau(>0)$ is the time the measurement finishes.
Here we adopt the following matrix element of $e^{\tau S}$:
\begin{equation}\label{ME1}
\left(
\begin{array}{cc}
a & b \\
c & d
\end{array}
\right)=e^{\tau S},
\end{equation}
which implies
\begin{equation}
\left(
\begin{array}{cc}
d & -c \\
-b & a
\end{array}
\right)=e^{-\tau S^T}
\end{equation}
since $ad-bc=1$.
The q-rms error $\varepsilon(Q_1)$ of $Q_1$ and the q-rms disturbance $\eta(P_1)$ of $P_1$ then are given by
\begin{subequations}
\begin{align}
\varepsilon(Q_1)^2 &=\langle \psi\otimes\xi| (Q_2(\tau)-Q_1(0))^2(\psi\otimes\xi)\rangle \nonumber\\
= & (c-1)^2\sigma(Q_1)^2+d^2\sigma(Q_2)^2+ ((c-1)\langle Q_1 \rangle+d\langle Q_2 \rangle)^2, \label{error} \\
\eta(P_1)^2 &=\langle \psi\otimes\xi| (P_1(\tau)-P_1(0))^2(\psi\otimes\xi)\rangle \nonumber \\
= & (d-1)^2\sigma(P_1)^2+c^2\sigma(P_2)^2+ ((d-1)\langle P_1 \rangle-c\langle P_2 \rangle)^2, \label{disturbance}
\end{align}
\end{subequations}
respectively. 
When using a linear position measurement $\mathbb{M}=(\mathcal{H}_\mathbf{P},\xi,Q_2,U(\tau))$ for $\mathbf{S}$,
$Q_1(0)$ and $Q_2(\tau)$ are mutually commuting, and so are $P_1(0)$ and $P_1(\tau)$,
so that
\begin{subequations}
\begin{align}
\varepsilon(Q_1,\mathbb{M},\psi) &=\varepsilon_G(\mu^{Q_1(0),Q_2(\tau)}_{\psi\otimes\xi}), \label{G11} \\
\eta(P_1,\mathbb{M},\psi)&=\varepsilon_G(\mu^{P_1(0),P_1(\tau)}_{\psi\otimes\xi}). \label{G21}
\end{align}
\end{subequations}

\section{Results \label{3}}

\subsection{Characterization theorem \label{3.1}}
\begin{figure}[H]
\begin{center}
\includegraphics[clip]{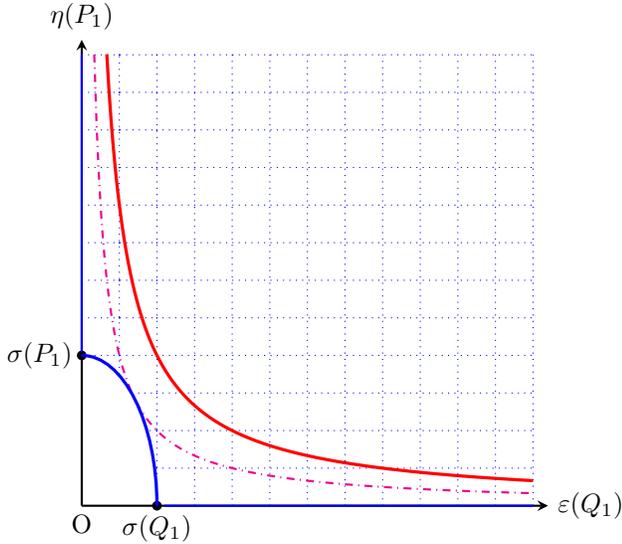}
\caption{When the state of $\mathbf{S}$ is $\psi$,
possible values of the pair $(\varepsilon(Q_1),\eta(P_1))$
are indicated by the area with a grid of dotted blue lines and with blue boundary
except for two points $(\sigma(Q_1),0)$ and $(0,\sigma(P_1))$. By the theorem, 
$\varepsilon(Q_1)^2\sigma(P_1)^2+\sigma(Q_1)^2\eta(P_1)^2=\hbar^2/4$
($\varepsilon(Q_1),\eta(P_1)>0$), a part of its boundary,
is achieved by linear position measurements,
and gives the unbreakable limitation for the pair $(\varepsilon(Q_1),\eta(P_1))$.
The red line is Heisenberg's bound, $\varepsilon(Q_1)\eta(P_1)=\hbar/2$.
On the other hand, the dashed magenta line indicates $\varepsilon(Q_1)\eta(P_1)=\hbar/4$.}\label{fig:}
\end{center}
\end{figure}

The first main result is summarized as the following theorem:
\begin{theorem}{}\label{Main1}
$\quad$\\
A linear position measurement $\mathbb{M}=(\mathcal{H}_\mathbf{P},\xi,Q_2,U(\tau))$ for $L^2(\mathbb{R})$
satisfies Eq.~(\ref{BOinq2}) in $\psi$ if and only if it satisfies the following two conditions:\\
$(i)$ $c>0$, $d>0$ and $c+d=1$.\\
$(ii)$ $\xi$ is equal to the minimum uncertainty state $\xi_c$ with $\langle Q_2 \rangle_{\xi_c}=q_1$,
$\langle P_2 \rangle_{\xi_c}=-p_1$ and $\sigma(Q_2\Vert\xi_c)=\sqrt{\frac{c}{1-c}}\sigma_1$, i.e.,
\begin{equation}
\xi_c(y)=\sqrt[4]{\dfrac{1-c}{(2\pi)c\sigma_1^2}}e^{-\frac{1-c}{4c\sigma_1^2}(y-q_1)^2-i\frac{p_1}{\hbar} y},
\hspace{5mm}y\in\mathbb{R},
\end{equation}
in the coordinate representation.\\
If a linear position measurement $\mathbb{M}$ has the minimum error-disturbance in $\psi$, then we have
\begin{equation}\label{MUED2}
\varepsilon(Q_1)^2=(1-c)\sigma(Q_1)^2\hspace{3mm}\text{and}\hspace{3mm}\eta(P_1)^2=c\sigma(P_1)^2.
\end{equation}

Furthermore, for every $\mu\in(0,1)$, there exists a linear position measurement
for $L^2(\mathbb{R})$ satisfying 
\begin{equation}\label{MUED}
\varepsilon(Q_1)^2=(1-\mu)\sigma(Q_1)^2\hspace{3mm}\text{and}\hspace{3mm}\eta(P_1)^2=\mu\sigma(P_1)^2
\end{equation}
in $\psi$.
\end{theorem}
By the above theorem, any linear position measurement with the minimum error-disturbance in $\psi$ satisfies
\begin{subequations}
\begin{align}
\varepsilon(Q_1) &=\sqrt{1-\mu}\sigma(Q_1)<\sigma(Q_1), \\
\eta(P_1) &= \sqrt{\mu} \sigma(P_1)<\sigma(P_1), \\
\varepsilon(Q_1)\eta(P_1)&=\dfrac{\hbar}{2}\sqrt{\dfrac{1}{4}-\left(\mu-\dfrac{1}{2} \right)^2}
\leq \dfrac{\hbar}{4}<\dfrac{\hbar}{2}.
\end{align}
\end{subequations}
The region of possible values of the pair $(\varepsilon(Q_1),\eta(P_1))$
is drawn in FIG. \ref{fig:}, when using position measurements in $\psi$.

\subsection{Linear position measurements with the minimum error-disturbance in $\psi$ \label{3.2}}

Next, we present three families
$\{\mathbb{A}_\mu\}_{\mu\in(0,1)}$, $\{\mathbb{B}_\mu\}_{\mu\in(0,1)}$ and $\{\mathbb{C}_\mu\}_{\mu\in(0,1)}$
of linear position measurements satisfying Eq.~(\ref{MUED}) in $\psi$ for each $\mu\in(0,1)$.
$\mathbb{A}_\mu$, $\mathbb{B}_\mu$ and $\mathbb{C}_\mu$, respectively,
are linear position measurements $(\mathcal{H}_\mathbf{P},\xi_\mu,Q_2,U(\tau))$
whose parameters $\tau$, $\alpha$, $\beta$ and $\gamma$ are given in Table \ref{t1}.
Here $D$ used in Table \ref{t1} is defined by $D=\det S=-(\gamma^2+\alpha\beta)$.
\begin{table}[H]
\begin{center}
\caption{}
\begin{ruledtabular}
\begin{tabular}{cccccc} 
 & $\tau$ & $\alpha$ & $\beta$ & $\gamma$ & $D$ \\ \hline 
$\mathbb{A}_\mu$ & $\arccos (1-\mu)$ & $\sqrt{\frac{\mu}{2-\mu}}$ & $-\sqrt{\frac{2-\mu}{\mu}}$ & $0$ & $1$ \\ \hline
$\mathbb{B}_\mu$ & $\mu$ & $1$ & $-1$ & $1$ & $0$ \\ \hline
$\mathbb{C}_\mu$ & $-\log(1-\mu)$  & $\frac{2(1-\mu)}{2-\mu}$ & $0$ & $1$ & $-1$ \\
\end{tabular} \label{t1}
\end{ruledtabular}
\end{center}
\end{table}
From Table \ref{t1}, for every $\mu\in(0,1)$, $e^{\tau S}$ and $e^{-\tau S^T}$ for
$\mathbb{A}_\mu$, $\mathbb{B}_\mu$, and $\mathbb{C}_\mu$, respectively, are then given in Table \ref{t2}.
\begin{table}[H] 
\begin{center}
\caption{}
\begin{ruledtabular}
\begin{tabular}{ccc} 
 & $e^{\tau S}$ & $e^{-\tau S^T}$  \\ \hline
$\mathbb{A}_\mu$ & $\left(
\begin{array}{cc}
1-\mu  & \mu-2 \\
\mu & 1-\mu
\end{array}
\right)$ & $\left(
\begin{array}{cc}
1-\mu & -\mu \\
2-\mu & 1-\mu
\end{array}
\right)$ \\ \hline
$\mathbb{B}_\mu$ & $\left(
\begin{array}{cc}
1+\mu & -\mu \\
\mu & 1-\mu
\end{array}
\right)$ & $\left(
\begin{array}{cc}
1-\mu & -\mu \\
\mu & 1+\mu
\end{array}
\right)$ \\ \hline
$\mathbb{C}_\mu$ & $\left(
\begin{array}{cc}
\frac{1}{1-\mu} & 0 \\
\mu & 1-\mu
\end{array}
\right)$ & $\left(
\begin{array}{cc}
1-\mu & -\mu \\
0 & \frac{1}{1-\mu}
\end{array}
\right)$ \\ 
\end{tabular} \label{t2}
\end{ruledtabular}
\end{center}
\end{table}
Let $\mu\in(0,1)$. $\mathbb{A}_\mu$, $\mathbb{B}_\mu$ and $\mathbb{C}_\mu$
satisfy the following relations:
\begin{subequations}
\begin{align}
&\hspace{3mm}\varepsilon(Q_1,\mathbb{A}_\mu,\psi)^2 =\varepsilon(Q_1,\mathbb{B}_\mu,\psi)^2 \nonumber \\
 &=\varepsilon(Q_1,\mathbb{C}_\mu,\psi)^2=(1-\mu)\sigma(Q_1)^2, \label{EModelABC1} \\
&\hspace{3mm}\eta(Q_1,\mathbb{A}_\mu,\psi)^2 =\eta(Q_1,\mathbb{B}_\mu,\psi)^2  \nonumber \\
 &=\eta(Q_1,\mathbb{C}_\mu,\psi)^2 =\mu \sigma(P_1)^2 \label{EModelABC2}
\end{align}
\end{subequations}

To quantitatively check the difference of $\{\mathbb{A}_\mu\}_{\mu\in(0,1)}$,
$\{\mathbb{B}_\mu\}_{\mu\in(0,1)}$ and $\{\mathbb{C}_\mu\}_{\mu\in(0,1)}$,
the q-rims disturbance $\eta(Q_1)$ of $Q_1$ is very useful.
Since every linear position measurement $\mathbb{M}$ with the minimum error-disturbance in $\psi$ satisfies
\begin{equation}
\eta(Q_1,\mathbb{M},\psi)^2=\left\{ (a-1)^2+b^2\dfrac{c}{1-c} \right\}\sigma_1^2+(a+b-1)^2q_1^2,
\end{equation}
we obtain
\begin{subequations}
\begin{align}
\lim_{\mu\rightarrow 1-0}\varepsilon(Q_1,\mathbb{A}_\mu,\psi)&\eta(Q_1,\mathbb{A}_\mu,\psi) = \sigma_1^2,\\
\lim_{\mu\rightarrow +0}\eta(Q_1,\mathbb{A}_\mu,\psi)&=2|q_1|, \\
\lim_{\mu\rightarrow 1-0}\varepsilon(Q_1,\mathbb{B}_\mu,\psi)&\eta(Q_1,\mathbb{B}_\mu,\psi) = \sigma_1^2,\\
\lim_{\mu\rightarrow +0}\eta(Q_1,\mathbb{B}_\mu,\psi)&=\lim_{\mu\rightarrow +0}\eta(Q_1,\mathbb{C}_\mu,\psi)=0,\\
\lim_{\mu\rightarrow 1-0}\varepsilon(Q_1,\mathbb{C}_\mu,\psi)&\eta(Q_1,\mathbb{C}_\mu,\psi) = +\infty
\end{align}
\end{subequations}
from Eq.~(\ref{EModelABC1}) and Table \ref{t2}.

\subsection{Probability distributions and families of posterior states \label{3.3}}
We have presented concrete linear position measurements with the minimum error-disturbance in $\psi$.
Our next interest is probability distributions and states after the measurement when using such measurements.

First, we calculate probability distributions related to $\varepsilon(Q_1)$ and $\eta(P_1)$.
Let $\mathbb{M}=(\mathcal{H}_\mathbf{P},\xi_c,Q_2,U(\tau))$ be a linear position measurement
with the minimum error-disturbance in $\psi$,
i.e., a linear position measurement satisfying the conditions in the theorem in $\psi$.
The probability density functions $p^{Q_1(0),Q_2(\tau)}_{\psi\otimes\xi_c}(x,y)$
and $p^{P_1(0),P_1(\tau)}_{\psi\otimes\xi_c}(z,w)$
of the joint probability distributions of $Q_1(0)$ and $Q_2(\tau)$ and of $P_1(0)$ and $P_1(\tau)$
in $\psi\otimes\xi_c$, respectively, are given as follows:
\begin{subequations}
\begin{align}
p^{Q_1(0),Q_2(\tau)}_{\psi\otimes\xi_c}(x,y) &= p_{(1-c)\sigma_1^2}(x-y)p_{c\sigma_1^2}(y-q_1), \label{JDQ1} \\
p^{P_1(0),P_1(\tau)}_{\psi\otimes\xi_c}(z,w) &= p_{c\hat{\sigma}_1^2}(z-w)p_{(1-c)\hat{\sigma}_1^2}(w-p_1), \label{JDQ2}
\end{align}
\end{subequations}
where $p_{\sigma^2}(x)$ denotes the probability density function of the Gaussian probability measure
with mean $0$ and variance $\sigma^2$, i.e.,
\begin{equation}
p_{\sigma^2}(x)=\dfrac{1}{\sqrt{(2\pi)\sigma^2}}e^{-\frac{1}{2\sigma^2}x^2}.
\end{equation}
We see that both depend on $\psi$ and $0<c<1$. From Eq.~(\ref{JDQ1}), we obtain
\begin{equation}
p^{Q_2(\tau)}_{\psi\otimes\xi_c}(y) = p_{c\sigma_1^2}(y-q_1) \label{QDF3}
\end{equation}
and, furthermore, the probability density of the conditional probability distribution of $Q_1(0)$
given the value $y$ of $Q_2(\tau)$ in $\psi\otimes\xi_c$ 
\begin{equation}
p^{Q_1(0)}_{Q_2(\tau)=y|\psi\otimes\xi_c}(x) = p_{(1-c)\sigma_1^2}(x-y).
\end{equation}
By Eq.~(\ref{MUED2}), we see that, given the value $y$ of $Q_2(\tau)$,
the difference of the value $x$ and $y$ of $Q_1(0)$ and $Q_2(\tau)$ obeys the Gaussian distribution with mean $0$
and standard deviation $\sqrt{1-c}\sigma_1=\varepsilon(Q_1)$.

Next, we consider family of posterior states, which is set of states after the measurement
for each output value of the meter (see \cite{ozawa1985,okamura2016measurement}
for the general theory of family of posterior states).
Let $\mathbb{M}$ be a linear position measurement with the minimum error-disturbance in $\psi$.
The family $\{\psi_y\}_{y\in\mathbb{R}}$ of posterior states for $(\mathbb{M},\psi)$ is the set of
the minimum uncertainty state $\psi_y$ with $\langle Q_1 \rangle_{\psi_y}=(a+b)y$,
$\langle P_1 \rangle_{\psi_y}=p_1$
and $\sigma(Q_1\Vert\psi_y)=\sigma_1/\sqrt{1-c}$ for all $y\in\mathbb{R}$, i.e., 
\begin{equation} \label{FPS}
\psi_y(x)=\sqrt[4]{\dfrac{1-c}{(2\pi)\sigma_1^2}}e^{-\frac{1-c}{4\sigma_1^2}(x-(a+b)y)^2
+i\frac{p_1}{\hbar} x}
\end{equation}
for all $y\in\mathbb{R}$ in the coordinate representation. For every interval $J$ of $\mathbb{R}$,
we then obtain the state $\rho_J$ after the measurement under the condition that
output values not contained in $J$ is excluded, which is given by
\begin{equation}
\mathrm{Tr}[X \rho_J]=\frac{\langle U(\tau)(\psi\otimes\xi_c)|X E^{Q_2}(J) U(\tau)(\psi\otimes\xi_c)\rangle}{
\langle U(\tau)(\psi\otimes\xi_c)| E^{Q_2}(J) U(\tau)(\psi\otimes\xi_c)\rangle}
\end{equation}
for all bounded operators $X$ on $\mathcal{H}_\mathbf{S}$, whenever
$\langle U(\tau)(\psi\otimes\xi_c)| E^{Q_2}(J) U(\tau)(\psi\otimes\xi_c)\rangle\neq 0$.
The family $\{\psi_y\}_{y\in\mathbb{R}}$ of posterior states for $(\mathbb{M},\psi)$ then satisfies
\begin{equation}
\rho_{J}=\dfrac{1}{\mu_{c \sigma_1^2,q_1}(J)}\int_J |\psi_y\rangle\langle\psi_y|\;
p_{c \sigma_1^2}(y-q_1)\;dy,
\end{equation}
where $\mu_{c \sigma_1^2,q_1}$ is the Gaussian measure with mean $q_1$ and variance $c\sigma_1^2$, i.e.,
$\mu_{c\sigma_1^2,q_1}(J)=\int_J p_{c \sigma_1^2}(y-q_1)dy$
for all intervals $J$ of $\mathbb{R}$(, more generally, for all Borel sets $J$ of $\mathbb{R}$).

\section{Discussion \label{4}}

The Branciard-Ozawa EDR is given by
\begin{align}
&\varepsilon(A)^2\sigma(B)^2+\sigma(A)^2\eta(B)^2 \nonumber\\
 &\hspace{3mm}+2\varepsilon(A)\eta(B)\sqrt{\sigma(A)^2\sigma(B)^2-D_{AB}^2}\geq D_{AB}^2
\end{align}
for observables $A$ and $B$.
Here $\sigma(A)^2=\mathrm{Tr}[A^2\rho]-(\mathrm{Tr}[A\rho])^2$,
$\sigma(B)^2=\mathrm{Tr}[B^2\rho]-(\mathrm{Tr}[B\rho])^2$ and
$D_{AB}=\mathrm{Tr}|\sqrt{\rho}(-i[A,B])\sqrt{\rho}|/2\geq
C_{AB}=|\mathrm{Tr}([A,B]\rho)|/2$, where $\rho$ is the density operator
describing a state of the system. For any vector state $\rho=|\phi\rangle\langle\phi|$,
$D_{AB}$ is equal to $C_{AB}$.
Eq.~(\ref{BOinq}) holds for the case where $A=Q_1$, $B=P_1$ and $\rho=|\psi\rangle\langle\psi|$.
Spin measurements that achieve the lower bound of the improved version of the inequality
in some class of states have already been constructed \cite{ozawa2014errordisturbance}.
On the other hand, the first achievement for
constructing position measurements with minimum error-disturbance in some class of states is made in the paper.

In the proof of Heisenberg's EDR
\begin{equation}
\varepsilon(Q_1)\eta(P_1)\geq \hbar/2
\end{equation}
by Heisenberg \cite{Heisenberg1927} and Kennard \cite{Kennard1927}, the approximate repeatability hypothesis (ARH),
the approximate version of the repeatable hypothesis (RH), is assumed.
In particular, Heisenberg only dealt with the case where the states after the measurement
are minimum uncertainty states.
The (A)RH is a natural assumption at that time, but is abandoned since the 1980s
(see \cite{ozawa2015heisenberg} for the details on  the RH and the ARH, and also
\cite{dirac1958principles,von2018mathematical,schrodinger1935gegenwartige,davies1970operational}).
Quantum measurement theory has changed significantly from what it was when quantum mechanics was established.


Following Laplace's pioneering investigation, Gauss \cite{Gauss1995theory} defined his rms error in 1821.
It is now widely used in statistics and experimental science.
Gauss' error is not always applicable to quantum systems
due to the noncommutativity of observables, but we should try it if applicable.
In other words, its universal validity is lost in quantum theory,
but it does not mean that it is always useless.
As given in Sec.~\ref{2.2}, its use is reasonable as long as linear position measurements are considered.
In particular, the noise-operator based q-rms error and Gauss's error coincide
when using linear position measurements. The author would like to emphasize that
the results of the paper \cite{ozawa1988measurement} and this paper show that
Heisenberg's EDR is already violated even when Gauss's error is available.
We must never miss this fact.

%


A q-rms error is said to be complete if it vanishes only for precise measurements of observables.
As shown in \cite{busch2004noise,ozawa2019soundness} for example,
the noise-operator based q-rms error is not complete.
On the other hand, it is recently verified in \cite{ozawa2019soundness} that the noise-operator based q-rms error satisfies
several satisfactory conditions except for the completeness from the operational point of view.
Furthermore, several complete q-rms errors, improvements of the noise-operator based q-rms error,
are also defined in \cite{ozawa2019soundness}.
Only the lack of the completeness does not mean that the noise-operator based q-rms error is useless.
In fact, it is widely applicable and useful enough as a standard.
The results of the paper also contribute to showing its value.
The author thinks that it is better to choose alternatives
when there is a problem with the use of the noise-operator based q-rms error.
To this end, it will be increasingly important to study complete q-rms errors as a candidate for alternatives.



\section{Methods \label{5}}

As in the standard textbook of quantum mechanics, for $j,k=1,2$, $Q_j$ and $P_k$ satisfy
\begin{align*}
(Q_j f)(x_1,x_2) &= x_j  f(x_1,x_2), \\
(P_k g)(x_1,x_2) &= \dfrac{\hbar}{i} \dfrac{\partial}{\partial x_k} g(x_1,x_2)
\end{align*}
in the coordinate representation for suitable functions $f$ and $g$ on $\mathbb{R}^2$.
We do not explicitly use these representations here.

\subsection{Proof of Theorem \label{5.1}}
Let $\mathbb{M}=(\mathcal{H}_\mathbf{P},\xi,Q_2,U(\tau))$ be a linear position measurement for $\mathcal{H}_\mathbf{S}$.
By Eqs.~(\ref{error}), (\ref{disturbance}), the following evaluation holds for $\mathbb{M}$:
\begin{align}
 &\hspace{5mm}\varepsilon(Q_1)^2\sigma(P_1)^2+\sigma(Q_1)^2\eta(P_1)^2 \nonumber\\
 &\geq (c-1)^2\sigma(Q_1)^2\sigma(P_1)^2+d^2\sigma(Q_2)^2\sigma(P_1)^2 \nonumber \\
 &\hspace{5mm}+(d-1)^2\sigma(Q_1)^2\sigma(P_1)^2+c^2\sigma(Q_1)^2\sigma(P_2)^2 \nonumber \\
 &=\dfrac{\hbar^2}{4}\{(c-1)^2+(d-1)^2\} \nonumber \\
 &\hspace{5mm} +c^2\sigma(Q_1)^2\sigma(P_2)^2+d^2\sigma(Q_2)^2\sigma(P_1)^2 \nonumber \\
 &=\dfrac{\hbar^2}{4}\{(c-1)^2+(d-1)^2\}+2|cd|\sigma(Q_1)\sigma(P_1)\sigma(Q_2)\sigma(P_2) \nonumber \\
 &\hspace{5mm} +(|c|\sigma(Q_1)\sigma(P_2)-|d|\sigma(Q_2)\sigma(P_1))^2 \nonumber \\
 &\geq \dfrac{\hbar^2}{4}\{(c-1)^2+(d-1)^2\}+\hbar|cd|\sigma(Q_2)\sigma(P_2) \nonumber \\
 &\geq \hbar^2 l(c,d),
\end{align}
where $l(c,d)$ is defined by
\begin{equation}
l(c,d)=\dfrac{1}{4}\{(c-1)^2+(d-1)^2\}+\dfrac{1}{2}|cd|,
\end{equation}
and takes the minimal value $1/4$ when $c,d\geq 0$ and $c+d=1$.
$\mathbb{M}$ has the minimum error-disturbance in $\psi$ if and only if it satisfies the following four conditions:\\
$(1)$ $(c-1)\langle Q_1 \rangle+d\langle Q_2 \rangle=0$ and
$(d-1)\langle P_1 \rangle-c\langle P_2 \rangle=0$.\\
$(2)$ $|c|\sigma(Q_1)\sigma(P_2)=|d|\sigma(Q_2)\sigma(P_1)$.\\
$(3)$ $\sigma(Q_2)\sigma(P_2)=\hbar/2$.\\
$(4)$ $c,d\geq 0$ and $c+d=1$.\\
$(3)$ implies that $\xi$ is a minimum uncertainty state.
By $\sigma(Q_1)\sigma(P_1)=\sigma(Q_2)\sigma(P_2)=\hbar/2$ and $(2)$,
\begin{equation}\label{SD1}
\sigma(Q_2)^2 = \dfrac{\hbar}{2} \dfrac{\sigma(Q_2)}{\sigma(P_2)}
 = \dfrac{\hbar}{2} \dfrac{|c|}{|d|} \dfrac{\sigma(Q_1)}{\sigma(P_1)}
 =\dfrac{|c|}{|d|}\sigma(Q_1)^2 
\end{equation}
This is consistent with the condition $(3)$ when $c,d\neq 0$.
Therefore, under the condition $(i)$ in the theorem, by Eq.~(\ref{SD1}) and the condition $(1)$,
$\xi$ is a minimum uncertainty state such that
$\langle Q_2 \rangle_\xi=q_1$, $\langle P_2 \rangle_\xi=-p_1$ and
$\sigma(Q_2\Vert\xi)=\sqrt{\dfrac{c}{1-c}}\sigma_1$, that is, $\xi=\xi_c$.
Under the condition $(i)$ in the theorem, $\xi=\xi_c$ implies the conditions $(1)$, $(2)$ and $(3)$.
Then we obtain
\begin{align*}
\varepsilon(Q_1)^2 &=(c-1)^2\sigma_1^2+(1-c)^2 \dfrac{c\sigma_1^2}{1-c}=(1-c)\sigma_1^2,\\
\eta(P_1)^2 &= c^2\hat{\sigma}_1^2+c^2\dfrac{1-c}{c}\hat{\sigma}_1^2=c\hat{\sigma}_1^2.
\end{align*}
This completes the proof of the first half of the theorem.

For every $\mu\in(0,1)$, we find linear position measurements such that $c=\mu$ and $d=1-\mu$.
To explicitly give linear position measurements satisfying Eq.~(\ref{MUED}) in $\psi$ for every $\mu\in(0,1)$,
we use a more explicit formula of $e^{tS}$:
\begin{equation}\label{ExpFor}
e^{tS}=\left\{
\begin{array}{ll}
\displaystyle{(\cos(t\sqrt{D}))I+\dfrac{\sin(t\sqrt{D})}{\sqrt{D}}S}, &\quad (D>0), \\
I+tS, &\quad (D=0), \\
\displaystyle{(\cosh(t\sqrt{-D}))I+\dfrac{\sinh(t\sqrt{-D})}{\sqrt{-D}}S}, &\quad (D<0).
\end{array}
\right.
\end{equation}
Since $c=\mu>0$, $\alpha$ must be non-zero, so that
$\beta$ is uniquely determined by
\begin{equation}
\beta=-\dfrac{\gamma^2+D}{\alpha}.
\end{equation}
From now on, we take the position that $\alpha$, $\gamma$, and $D$ are the fundamental variables,
and $\beta$ is determined from them.
Then, the cases $D>0$, $D=0$, $D<0$ must be handled separately.

[$D>0$] By Eqs.~(\ref{ME1}), (\ref{ExpFor}), both $c=\mu$ and $d=1-\mu$ become
\begin{equation}
\left\{
\begin{array}{l}
\displaystyle{\mu=\dfrac{\alpha}{\sqrt{D}}\sin(\tau\sqrt{D})}. \\
\displaystyle{1-\mu=\cos(\tau\sqrt{D})-\dfrac{\gamma}{\sqrt{D}}\sin(\tau\sqrt{D})}. 
\end{array}
\right.
\end{equation}
This system of equations is equivalent to 
\begin{equation} \label{D>0}
\left\{
\begin{array}{ll}
\cos(\tau\sqrt{D})=\dfrac{\gamma}{\alpha}\mu+(1-\mu),&  \\
\sin(\tau\sqrt{D})=\dfrac{\sqrt{D}}{\alpha}\mu. &
\end{array}
\right.
\end{equation}
For every $0<\mu<1$, $\gamma\geq 0$ and $D>0$, we have
$\dfrac{\gamma}{\alpha}\mu+(1-\mu)>0$ and $\dfrac{\sqrt{D}}{\alpha}\mu>0$ for all $\alpha>0$. 
Moreover, the function 
\begin{equation*}
u_+(\alpha)=\left\{\dfrac{\gamma}{\alpha}\mu+(1-\mu)\right\}^2
+\left(\dfrac{\sqrt{D}}{\alpha}\mu \right)^2
\end{equation*}
on $\mathbb{R}_+=\{\alpha\in\mathbb{R}\;|\;\alpha>0\}$ is monotone decreasing
and satisfies $\lim_{\alpha\rightarrow +0}u_+(\alpha)=+\infty$ and
$\lim_{\alpha\rightarrow +\infty}u_+(\alpha)=(1-\mu)^2<1$,
so that there uniquely exists $\alpha>0$ such that $u_+(\alpha)=1$.
There then uniquely exists $0<\tau<\dfrac{\pi}{2\sqrt{D}}$ satisfying Eq.~(\ref{D>0}).
For every $0<\mu<1$, $D>0$ and $\gamma\geq 0$,
there uniquely exist $\alpha>0$ and $0<\tau<\dfrac{\pi}{2\sqrt{D}}$
such that Eq.~(\ref{D>0}). For every $\mu\in(0,1)$, the model $\mathbb{A}_\mu$ is included in this case.
\textit{This completes the proof of the theorem.}

[$D=0$] By Eqs.~(\ref{ME1}), (\ref{ExpFor}), both $c=\mu$ and $d=1-\mu$ become
\begin{equation}
\left\{
\begin{array}{l}
\displaystyle{\mu=\alpha \tau} \\
\displaystyle{1-\mu=1-\gamma\tau}, 
\end{array}
\right.
\end{equation}
This system of equations and $D=0$ then imply 
\begin{equation}
\alpha=-\beta=\gamma=\dfrac{\mu}{\tau}. \label{D=0}
\end{equation}
For every $0<\mu<1$ and $\gamma>0$, there uniquely exist $\alpha>0$, $\beta<0$ and $\tau>0$
such that Eq.~(\ref{D=0}).
For every $\mu\in(0,1)$, the model $\mathbb{B}_\mu$ is included in this case.

[$D<0$] By Eqs.~(\ref{ME1}), (\ref{ExpFor}), both $c=\mu$ and $d=1-\mu$ become
\begin{equation}
\left\{
\begin{array}{l}
\displaystyle{\mu=\dfrac{\alpha}{\sqrt{-D}}\sinh(\tau\sqrt{-D})}, \\
\displaystyle{1-\mu=\cosh(\tau\sqrt{-D}) -\dfrac{\gamma}{\sqrt{-D}}\sinh(\tau\sqrt{-D})},
\end{array}
\right.
\end{equation}
so that $\alpha\neq 0$. This system of equations is equivalent to
\begin{equation}\label{D<0}
\left\{
\begin{array}{ll}
\cosh(\tau\sqrt{-D})=\dfrac{\gamma}{\alpha}\mu+(1-\mu), &  \\
\sinh(\tau\sqrt{-D})=\dfrac{\sqrt{-D}}{\alpha}\mu. &\quad
\end{array}
\right.
\end{equation}
Here we assume that $\alpha>0$ and $\gamma>0$.
The function $u_-$ on $\mathbb{R}_+$ is then defined
by 
\begin{equation}
u_-(\alpha)=\left\{\dfrac{\gamma}{\alpha}\mu+(1-\mu)\right\}^2-\left(\dfrac{\sqrt{-D}}{\alpha}\mu \right)^2.
\end{equation}
There exists a solution $\alpha>0$ of $u_-(\alpha)=1$ if and only if
\begin{equation}
(2-\mu)\alpha^2-2(1-\mu)\gamma\alpha-\mu(\gamma^2+D)=0.
\end{equation}
Since $1-\mu>0$, $\gamma>0$ and $1-(1-\mu)^2>0$ are always satisfied,
the above quadratic equation has a unique solution $\alpha>0$ only when $-\mu(\gamma^2+D)\leq 0$.
Thus, $\gamma$ must satisfy $\gamma\geq\sqrt{-D}$. 
For every $0<\mu<1$, $D<0$ and $\gamma\geq \sqrt{-D}$, there uniquely exist
$\alpha>0$ and $\tau>0$ such that Eq.~(\ref{D<0}).
For every $\mu\in(0,1)$, the model $\mathbb{C}_\mu$ is included in this case.

\subsection{Probability distributions and families of posterior states \label{5.2}}

The characteristic function $\lambda$ of the probability measure $\mu$ on $\mathbb{R}^d$ is defined as
the inverse Fourier transform of $\mu$: For every $k\in\mathbb{R}^d$,
\begin{equation}
\lambda(k)=\int_{\mathbb{R}^d} e^{i\langle x, k\rangle}\;d\mu(x),
\end{equation}
where $\langle \cdot,\cdot \rangle$ is the inner product on $\mathbb{R}^d$.
A probability measure on $\mathbb{R}^d$ is a Gaussian measure (also called a multivariate normal distribution)
\begin{equation}
\mu_{V,m}(dx)=\dfrac{1}{\sqrt{(2\pi)^d\det(V)}}e^{-\frac{1}{2}\langle x-m,V^{-1} (x-m)\rangle}\;dx
\end{equation}
if and only if its characteristic function has the form
\begin{equation}
\lambda_{V,m}(k)=e^{i\langle m,k \rangle-\frac{1}{2} \langle k,Vk \rangle},\hspace{5mm}k\in\mathbb{R}^d,
\end{equation}
where $V>0$ is a covariance matrix, $m\in\mathbb{R}^d$ is a mean vector.
For more on this basic fact, see the section on
characteristic functions in the standard textbook of probability theory and statistics.

Let $\mathbb{M}=(\mathcal{H}_\mathbf{P},\xi_c,Q_2,U(\tau))$
be a linear position measurement with the minimum error-disturbance in $\psi$.
For every $k=\left(
\begin{array}{c}
k_1  \\
k_2
\end{array}
\right)\in\mathbb{R}^2$, the characteristic function of $\mu^{Q_1(0),Q_2(\tau)}_{\psi\otimes\xi_c}$ is given by
\begin{align}
\lambda^{Q_1(0),Q_2(\tau)}_{\psi\otimes\xi_c}(k)
 &=\langle \psi\otimes\xi_c| e^{ik_1Q_1(0)+ik_2Q_2(\tau)} (\psi\otimes\xi_c) \rangle \nonumber \\
 &=\langle \psi|e^{i[k_1+ck_2]Q_1} \psi \rangle \langle \xi_c|e^{i[dk_2]Q_2} \xi_c \rangle \nonumber \\
 &=e^{iq_1(k_1+ck_2)-\frac{1}{2}\sigma_1^2(k_1+ck_2)^2}\nonumber \\
 &\hspace{5mm}\times e^{iq_1(dk_2)-\frac{1}{2}\frac{c}{1-c}\sigma_1^2(dk_2)^2}\nonumber \\
 &= e^{iq_1k_1+iq_1k_2-\frac{1}{2}\langle k,W k\rangle},
\end{align}
where $W=\sigma_1^2\left(
\begin{array}{cc}
1 & c \\
c & c
\end{array}
\right)$. We obtain Eq.~(\ref{JDQ1}) from $\det(W) = \sigma_1^4c(1-c)$ and
\begin{equation}
W^{-1} = \dfrac{1}{(1-c)\sigma_1^2}\left(
\begin{array}{cc}
1 & -1 \\
-1 & 1
\end{array}
\right)+\dfrac{1}{c\sigma_1^2}\left(
\begin{array}{cc}
0 & 0 \\
0 & 1
\end{array}
\right).
\end{equation}
Eq.~(\ref{JDQ2}) is derived in the same way.

Let $\mathbb{M}=(\mathcal{H}_\mathbf{P},\xi_c,Q_2,U(\tau))$ be
a linear position measurement with the minimum error-disturbance in $\psi$.
In order to find family of posterior states for $(\mathbb{M},\psi)$,
we also check the following probability density functions via their characteristic functions:
\begin{subequations}
\begin{align}
p^{Q_1(\tau),Q_2(\tau)}_{\psi\otimes\xi_c}(x,y) &= p_{\frac{\sigma_1^2}{1-c}}(x-(a+b)y)p_{c\sigma_1^2}(y-q_1),\label{QDF1}\\
p^{P_1(\tau),Q_2(\tau)}_{\psi\otimes\xi_c}(z,y) 
&= p_{(1-c)\hat{\sigma}_1^2}(z-p_1)p_{c\sigma_1^2}(y-q_1). \label{QDF2}
\end{align}
\end{subequations}
The formula $e^{iuQ_j+ivP_j}=e^{i\frac{\hbar}{2}uv}e^{iuQ_j}e^{ivP_j}$ for all $u,v\in\mathbb{R}$ and $j=1,2$
is applied to calculate the characteristic functions correponding to the probability distributions
that have the above densities.
By Eqs.~(\ref{QDF1}), (\ref{QDF2}) and (\ref{QDF3}),
the density functions $p^{Q_1(\tau)}_{Q_2(\tau)=y|\psi\otimes\xi_c}(x)$ and 
$p^{P_1(\tau)}_{Q_2(\tau)=y|\psi\otimes\xi_c}(z)$ of the conditional probability distributions of
$Q_1(\tau)$ and $P_1(\tau)$ given the value $y$ of $Q_2(\tau)$ in $\psi\otimes\xi_c$ are given by
\begin{subequations}
\begin{align}
p^{Q_1(\tau)}_{Q_2(\tau)=y|\psi\otimes\xi_c}(x) &= p_{\frac{\sigma_1^2}{1-c}}(x-(a+b)y),\\
p^{P_1(\tau)}_{Q_2(\tau)=y|\psi\otimes\xi_c}(z)
&= p_{(1-c)\hat{\sigma}_1^2}(z-p_1),
\end{align}
\end{subequations}
respectively. The relation $\dfrac{\sigma_1^2}{1-c}\cdot (1-c)\hat{\sigma}_1^2=\dfrac{\hbar^2}{4}$
implies that the family of posterior states for $(\mathbb{M},\psi)$ is given by Eq.~(\ref{FPS})
 and is unique up to phase.

\section{Summary and perspective \label{6}}

In the paper, linear position measurements with the minimum error-disturbance
in each minimum uncertainty state have been constructed for each possible error value.
This is nothing but the result showing that the lower bound of the Branciard-Ozawa inequality
for the q-rms error of position and the q-rms disturbance of momentum in each minimum uncertainty state is achievable.
It is also the first achievement for position measurements
since the reformulation of uncertainty relations has started.
Moreover, we have obtained joint probability distributions and families of posterior states when using them.
In order to quantitatively examine the effect of the measurement,
the q-rms disturbance of position has also been analyzed.
It is expected to construct measurements with minimum error-disturbance
in a broader class of states in the future,
which will lead to a new understanding of quantum limits, including uncertainty relations.
Linear simultaneous measurements of position and momentum with minimum error-trade-off
in each minimum uncertainty state will be discussed in the subsequent paper \cite{okamura2020minimaluncertainty2}.

\begin{acknowledgments}
The author thanks Prof. Motoichi Ohtsu and Prof. Fumio Hiroshima for their warmful encouragements.
He also thanks Prof. Masanao Ozawa for useful comments.
\end{acknowledgments}


\bibliography{sample}

\end{document}